\documentclass[twocolumn]{aastex631}

\usepackage{amsmath, amstext, verbatim, graphicx, xcolor, longtable}

\newcommand{\Hb}{\hbox{{\rm H}$\beta$}}
\newcommand{\Ha}{\hbox{{\rm H}$\alpha$}}
\newcommand{\NII}{\hbox{{\rm [N}\kern 0.1em{\sc ii}{\rm ]}}}
\newcommand{\OIII}{\hbox{{\rm [O}\kern 0.1em{\sc iii}{\rm ]}}}

\begin{document}
\title{\large \bf Here There Be (Dusty) Monsters: High Redshift AGN are Dustier Than Their Hosts
}
\shorttitle{Dusty Monsters}
\shortauthors{Brooks et al.}
\


\author[0000-0001-5384-3616]{Madisyn Brooks}
\affil{Department of Physics, 196A Auditorium Road, Unit 3046, University of Connecticut, Storrs, CT 06269, USA}

\author[0000-0002-6386-7299]{Raymond C.\ Simons}
\affiliation{Department of Engineering and Physics, Providence College, 1 Cunningham Sq, Providence, RI 02918 USA}
\affil{Department of Physics, 196A Auditorium Road, Unit 3046, University of Connecticut, Storrs, CT 06269, USA}

\author[0000-0002-1410-0470]{Jonathan R. Trump}
\affil{Department of Physics, 196A Auditorium Road, Unit 3046, University of Connecticut, Storrs, CT 06269, USA}

\author[0000-0003-1282-7454]{Anthony J. Taylor}
\affiliation{Department of Astronomy, The University of Texas at Austin, Austin, TX, USA}

\author[0000-0001-8534-7502]{Bren Backhaus}
\affiliation{Department of Physics and Astronomy, University of Kansas, Lawrence, KS 66045, USA}

\author[0000-0001-8047-8351]{Kelcey Davis}
\affiliation{Department of Physics, 196A Auditorium Road, Unit 3046, University of Connecticut, Storrs, CT 06269, USA}


\author{V\'eronique Buat}
\affiliation{Aix Marseille Univ, CNRS, CNES, LAM, Marseille, France}
\affiliation{Institut Universitaire de France (IUF), Paris, France}

\author[0000-0001-7151-009X]{Nikko J. Cleri}
\affiliation{Department of Astronomy and Astrophysics, The Pennsylvania State University, University Park, PA 16802, USA}
\affiliation{Institute for Computational and Data Sciences, The Pennsylvania State University, University Park, PA 16802, USA}
\affiliation{Institute for Gravitation and the Cosmos, The Pennsylvania State University, University Park, PA 16802, USA}

\author[0000-0002-6219-5558]{Alexander de la Vega}
\affiliation{Department of Physics and Astronomy, University of California, 900 University Ave, Riverside, CA 92521, USA}

\author[0000-0001-8519-1130]{Steven L. Finkelstein}
\affiliation{Department of Astronomy, The University of Texas at Austin, Austin, TX, USA}
\affiliation{Cosmic Frontier Center, The University of Texas at Austin, Austin, TX, USA}

\author[0000-0002-3301-3321]{Michaela Hirschmann}
\affiliation{Institute of Physics, Laboratory of Galaxy Evolution, Ecole Polytechnique Fédérale de Lausanne (EPFL), Observatoire de Sauverny, 1290 Versoix, Switzerland}

\author[0000-0002-4884-6756]{Benne W. Holwerda}
\affiliation{Department of Physics, University of Louisville, Natural Science Building 102, 40292 KY Louisville, USA}

\author[0000-0002-8360-3880]{Dale D. Kocevski}
\affiliation{Department of Physics and Astronomy, Colby College, Waterville, ME 04901, USA}

\author[0000-0002-6610-2048]{Anton M. Koekemoer}
\affiliation{Space Telescope Science Institute, 3700 San Martin Drive, Baltimore, MD 21218, USA}

\author[0000-0003-1581-7825]{Ray A. Lucas}
\affiliation{Space Telescope Science Institute, 3700 San Martin Drive, Baltimore, MD 21218, USA}

\author[0000-0001-9879-7780]{Fabio Pacucci}
\affiliation{Center for Astrophysics $\vert$ Harvard \& Smithsonian, 60 Garden St, Cambridge, MA 02138, USA}
\affiliation{Black Hole Initiative, Harvard University, 20 Garden St, Cambridge, MA 02138, USA}

\author[0000-0001-7755-4755]{Lise-Marie Seillé}
\affiliation{Aix Marseille Univ, CNRS, CNES, LAM, Marseille, France}

\begin{abstract}
\textit{JWST} spectroscopy has discovered a population of $z \gtrsim 3.5$ galaxies with broad Balmer emission lines, and narrow forbidden lines, that are consistent with hosting active galactic nuclei (AGN). Many of these systems, now known as ``little red dots" (LRDs), are compact and have unique colors that are very red in the optical/near-infrared and blue in the ultraviolet.  The relative contribution of galaxy starlight and AGN to these systems remains uncertain, especially for the galaxies with unusual blue+red spectral energy distributions.  In this work, we use Balmer decrements to measure the independent dust attenuation of the broad and narrow emission-line components of a sample of 29 broad-line AGN identified from three public JWST spectroscopy surveys: CEERS, JADES, and RUBIES. Stacking the narrow components from the spectra of 25 sources with broad \Ha\ and no broad \Hb\ results in a median narrow \Ha/\Hb\ = $2.47^{+0.05}_{-0.05}$ (consistent with $A_{v} = 0$) and broad \Ha/\Hb\ $> 8.85$ ($A_{v} > 3.63$). The narrow and broad Balmer decrements imply little-to-no attenuation of the narrow emission lines, which are consistent with being powered by star formation and located on larger physical scales. Meanwhile, the lower limit in broad \Ha/\Hb\ decrement, with broad \Hb\ undetected in the stacked spectrum of 25 broad-\Ha\ AGN, implies significant dust attenuation of the broad-line emitting region that is presumably associated with the central AGN. Our results indicate that these systems, on average, are consistent with heavily dust-attenuated AGN powering the red parts of their SED while their blue UV emission is powered by unattenuated star formation in the host galaxy.

\end{abstract}

\section{Introduction}\label{Introduction}
\textit{JWST} observations of extragalactic deep fields have unveiled a new regime of black hole (BH) science; faint, high-redshift ($z \gtrsim 3.5$) active galactic nuclei (AGN) are being detected in abundance through broad Balmer line emission and are consistent with being powered by BHs with inferred masses $10^{6} - 10^{8}$ $\rm{M_{\odot}}$ \citep{Kocevski2023, Larson2023, Harikane2023, Maiolino2023, Ubler2023, Killi2023, Kokorev2023, Greene2023, Kocevski2024, Taylor2024}. This sample of AGN allows us to probe the low-mass distribution of BHs at high-redshifts, providing insight into the first black holes, i.e., the population of black hole seeds \citep[e.g.,][]{Volonteri2003, BrommLoeb2003,Lodato2006, Inayoshi2020, Pacucci2023}.

\par A fraction ($\sim 20 \%$) of broad-line (BL) identified AGN are compact sources that appear heavily obscured and are characterized by a ``V-shaped" spectral energy distribution (SED) with a steep red continuum in the rest-frame optical and elevated blue colors in the UV \citep{Mathee2023, Greene2023, Kocevski2024, Barro2024}. The emission mechanisms that power the red+blue colors in these sources, now colloquially known as ``little red dots" (LRDs) \citep{Mathee2023}, have been heavily debated in the literature. The excess of UV light can be explained by light scattered from a central AGN or from an unobscured host galaxy, while the optical colors could be a dust-reddened AGN or emission from starburst galaxies \citep{Kocevski2023, Labbe2023, Barro2024, Li2024, Akins2024}. \text{JWST}-detected broad-line AGN are more abundant \citep{Harikane2023a, Harikane2023,Kokorev2023, Greene2024}, about 1-2 dex higher in number density, than what is expected from local quasar studies \citep{Vestergaard2009, Kelly2013,Matsuoka2018} and provide a unique population to further explore the $M_{BH}-M_{*}$ relation in the early Universe \citep{Pacucci2023, Durodola2024}. 

\par Notably, X-ray emission studies of LRDs have reported non-detections \citep{Lyu2024, Matthee2024, Ananna204}, and even stacking techniques still fail to produce an X-ray detection \citep{Yue2024, Maiolino2024} of these high redshift sources. The failure to detect BLAGN and/or LRDs in X-ray observations could be explained by X-ray absorption by gas in the BL region with large covering factors \citep{Maiolino2024} and/or super-Eddington accretion onto slowly spinning BHs, a combination that leads to intrinsically-weak SEDs in the X-rays \citep{Juodzbalis2024, PacucciandNarayan2024, Lambrides2024}. Analysis of two X-ray confirmed AGN, at $z = 3.1$ and $z = 4.66$ in \cite{Wang2024} and \cite{Kocevski2024}, find that their X-ray emission is consistent with a dust-reddened AGN and that they could be lower redshift analogs of the LRD population. 

\par In this paper, we investigate the optical dust attenuation of high-redshift ($z > 3.5$) BLAGN identified in the JWST deep fields. Our sample consists of 29 spectroscopically confirmed BLAGN detected through broad \Ha\ emission and is gathered from three large public spectroscopy surveys: CEERS, JADES, and RUBIES. We derive narrow-line and broad-line Balmer decrements, using \Ha\ and \Hb, for this sample of BLAGN to explore different physical scenarios that can contribute to the ``V-shaped" SEDs associated with LRDs. We stack sources that show no broad \Hb\ emission (25/29) to further constrain the dust attenuation seen in the narrow and broad line emission.

\par This paper is presented as follows. In \S \ref{Dataset} we describe our AGN sample, in \S \ref{Line Fitting} we describe our line fitting technique and Balmer decrement measurements, and in \S\ref{results} we describe our results and their implications for BLAGN. For this work, we assume a flat $\mathrm{\Lambda}$CDM cosmology with $\mathrm{H_{0}}$ = 67.4 km s $^{-1}$ Mpc$^{-1}$ and $\Omega_{\mathrm{M}} = 0.315$ \citep{Planck2020}
.


\section{Observational Dataset}\label{Dataset}
\begin{figure}[t!]
    \centering
    \includegraphics[width=\columnwidth]{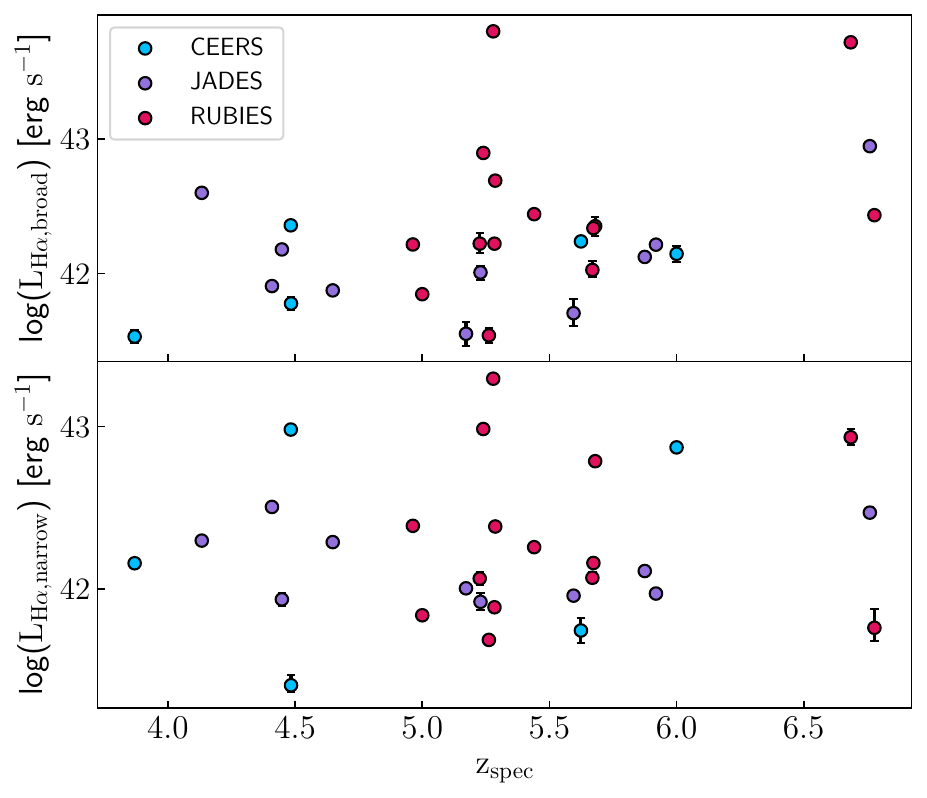}
    \caption{Broad-line component \Ha\ luminosity (top) and narrow-line component \Ha\ luminosity (bottom) as a function of redshift for our sample of 27 BLAGN, spanning $4.13 < z < 6.78$. CEERS sources are shown in blue, JADES sources are shown in purple, and RUBIES are shown in red.}
    \label{fig:sample luminosity}
\end{figure}

We select AGN in the JWST deep fields which exhibit broad \Ha\ emission and have both \Ha\ and \Hb\ spectral coverage  \citep{Harikane2023, Kocevski2023, Maiolino2023, Taylor2024}. We analyzed data from the CEERS (\cite{Finkelstein2023}, JADES \citep{JADESDR3}, and RUBIES \citep{deGraaff2024} surveys. Our sample consists of galaxies observed with NIRSpec medium resolution ( $\lambda / \Delta \lambda$ $\sim$1000) spectra. For this study, we do not use PRISM spectra; PRISM spectra have strongly varying wavelength-dependent spectral resolutions that can make it challenging to resolve emission (broad+narrow) on the bluer end of the spectrum, like \Hb. The complete sample studied in this paper spans the redshift range of $ 4.13 < z < 6.76 $. This redshift range is chosen to ensure coverage of both \Ha\ and \Hb\ in the medium-grating observations. In total the sample contains 29 sources: 5 observed through CEERS, 10 through JADES, and 14 through RUBIES. A brief description of the observation programs used in this study follows and our sample selection is described in \S \ref{Sample Selection}. Our complete sample of BLAGN is shown in Table \ref{AGN Sample Table} and the distribution of \Ha\ luminosities is shown in Figure \ref{fig:sample luminosity}.

\subsection{CEERS}
The Cosmic Evolution Early Release Science Survey (CEERS) covered $\sim$100 $\rm{arcmin^2}$ of the Extended Groth Strip (EGS). Six NIRSpec pointings were observed with the G140M/F100LP, G235M/F170LP, and G395M/F290LP grating/filter pairs resulting in a complete wavelength coverage from $\sim1-5$ $\mu$m. Each NIRSpec pointing was observed for 0.86~hr in each grating. For this study, we use only the G235M/170LP and G395M/F290LP grating/filter pairs. These two configurations have coverage of both \Ha\ and \Hb\ over a redshift range of $2.41 < z < 6.77$. The spectroscopic data were processed with the STScI JWST Calibration Pipeline version v1.8.5. We refer to \cite{ArrabalHaro2023} for a full description of the NIRSpec data reduction for CEERS. 

\subsection{JADES}
The JWST Advanced Deep Extragalactic Survey (JADES) covered  $\sim$175 $\rm{arcmin^2}$ in the GOODS-S and GOODS-N fields. We use NIRSpec data publicly released as part of JADES DR3 \citep{JADESDR3} and focus only on the G235M/170LP and G395M/F290LP grating/filter pairs. The JADES spectroscopic data was processed with a custom pipeline described in \cite{Ferruit2022}. JADES observations were split into three visits, with individual objects being observed with one, two, or three visits at 2.3~hr exposure time per visit. Objects observed in each visit reached up to $\sim$7~hr of exposure time. 
We refer to \cite{Eisenstein2023}, \cite{Bunker2023}, and \cite{JADESDR3} for a full description of the JADES survey and NIRSpec data reduction.

\subsection{RUBIES}
RUBIES observed 6 pointings in the CEERS (EGS) field and 6 pointings in the PRIMER-UDS Field \citep{deGraaff2024}. The spectroscopic data was processed with the STScI JWST Calibration Pipeline version 1.13.4. RUBIES observed with the G395M/F290LP grating/filter pair, giving us our highest redshift sources. Sources observed only with the G395M grating that cover both the \Ha\ and \Hb\ wavelength range will fall within redshifts $4.90 < z < 6.77$. Each pointing in this survey had an exposure time of 0.80 hrs. We refer to \cite{Taylor2024} for a full description of the RUBIES NIRSpec data reduction used here.

\subsection{BLAGN Sample}\label{Sample Selection}
Our sample consists of AGN identified from broad \Ha\ emission and is collected from \cite{Harikane2023a, Kocevski2023, Maiolino2023} and \cite{Taylor2024}. We note that the full BLAGN sample in the respective papers might not be represented in this sample. We select sources with significantly detected ($>3\sigma$) broad \Ha\ emission and sufficient ($\pm0.05~\rm{\mu m}$) spectral coverage around the \Hb\ emission line. We now briefly describe the detection method used by each study but we refer to the respective papers for a full description.

\subsubsection{CEERS BLAGN}
The sources in CEERS are described in detail in \cite{Kocevski2023}, \cite{Harikane2023}, and \cite{Taylor2024}. The two BLAGN identified in \cite{Kocevski2023} are also identified in \cite{Harikane2023}. \cite{Harikane2023} selects sources that have a broad ($\rm{FWHM > 1000\,km\, s^{-1}}$) for either \Ha\ or \Hb\ and narrow ($\rm{FWHM < 700\,km\, s^{-1}}$) forbidden \OIII\ and \NII\ emission lines. \cite{Taylor2024} selects sources with broad \Ha\ that have a $\rm{FWHM > 700\,km\, s^{-1}}$, the broad component is detected with S/N $>4$, and the redshift of the object is determined by at least three strong emission lines.

\subsubsection{JADES BLAGN}
The sources in JADES are described in detail in \cite{Maiolino2023}. \cite{Maiolino2023} selects sources that have a broad component in \Ha\ or \Hb\ without a broad component in the forbidden \OIII5007 line. Additionally, they require the broad component of the Balmer lines to be at least a factor of two broader than the narrow component and have a significance of at least $5\sigma$. The difference in the Bayesian Information Criterion between a model with only a narrow line and one with both narrow and broad limes must be greater than 6, i.e., $BIC_{narrow} - BIC_{broad+narrow} > 6.$

\subsubsection{RUBIES BLAGN}
The sources in RUBIES are described in detail in \cite{Taylor2024}. \cite{Taylor2024} selects sources with broad \Ha\ that have a $\rm{FWHM > 700\,km\, s^{-1}}$, the broad component is detected with S/N $>4$, and the redshift of the object is determined by at least three strong emission lines. 

\begin{deluxetable}{cccc}
\label{AGN Sample Table}
\tablecaption{BL AGN Sample}

\tablehead{\colhead{Name} & \colhead{R.A.} & \colhead{DEC} &
\colhead{$z_{\mathrm{spec}}$} \\ \colhead{} & \colhead{[deg]} & \colhead{[deg]} & \colhead{}}

\startdata
CEERS-11728 & 215.084870 & 52.970738 & 3.869 \\
JADES-GN-73488 & 189.197396 & 62.177233 & 4.133 \\
JADES-GN-11836 & 189.220587 & 62.263675 & 4.409 \\
JADES-GN-53757 & 189.269778 & 62.194208 & 4.448 \\
CEERS-1665 & 215.178197 & 53.059349 & 4.483 \\
CEERS-1236 & 215.145291 & 52.967291 & 4.484 \\
JADES-GS-8083 & 53.132846 & -27.801860 & 4.648 \\
RUBIES-EGS-46985 & 214.805654 & 52.809497 & 4.963 \\
RUBIES-EGS-17416 & 214.949482 & 52.845415 & 5.000 \\
JADES-GN-62309 & 189.248977 & 62.218350 & 5.172 \\
RUBIES-EGS-17301 & 214.987485 & 52.873115 & 5.226 \\
JADES-GN-77652 & 189.293228 & 62.199003 & 5.229 \\
RUBIES-EGS-50052* & 214.823454 & 52.830277 & 5.240 \\
RUBIES-EGS-13872 & 215.132933 & 52.970705 & 5.262 \\
RUBIES-EGS-42046 & 214.795368 & 52.788847 & 5.279 \\
RUBIES-EGS-60935 & 214.923373 & 52.925593 & 5.287 \\
RUBIES-EGS-926125 & 215.137081 & 52.988554 & 5.284 \\
CEERS-746 & 214.809145 & 52.868483 & 5.624 \\
JADES-GN-1093 & 189.179742 & 62.224628 & 5.595 \\
RUBIES-UDS-29813 & 34.453355 & -5.270717 & 5.440 \\
RUBIES-UDS-19521 & 34.383672 & -5.287732 & 5.669 \\
RUBIES-UDS-47509 & 34.264602 & -5.232586 & 5.673 \\
RUBIES-EGS-27915 & 214.844229 & 52.789595 & 5.680 \\ 
JADES-GN-61888 & 189.168016 & 62.217013 & 5.875 \\
JADES-GS-10013704 & 53.126535 & -27.818092 & 5.919 \\
CEERS-397 & 214.836183 & 52.882678 & 6.000 \\
RUBIES-EGS-49140 & 214.892248 & 52.877410 & 6.685 \\
JADES-GN-954 & 189.151966 & 62.259635 & 6.760 \\
RUBIES-UDS-807469 & 34.376139 & -5.310366 & 6.778 \\
\enddata
\tablecomments{*RUBIES-EGS-50052 is the same source as CEERS-02782 reported in \cite{Harikane2023}. We elected to use RUBIES-EGS-50052 in this paper due to its higher SNR as noted by \cite{Taylor2024}.}

\end{deluxetable}

\section{Emission Line Fitting}\label{Line Fitting}
We fit the \Ha\ and \Hb\ Balmer lines in each spectrum with a dual-component, narrow+broad, Gaussian model using the Markov Chain Monte Carlo (MCMC) routine from the Python \texttt{emcee} package \cite{emcee}. The dual component Gaussian model, adapting the methodology used in \cite{Larson2023} is:

\begin{equation}
\begin{split}
    f(\lambda)_{dual} = f_{c} + f_{nar} \exp{\bigg(-\frac{1}{2}\frac{(\lambda - \lambda_{0})}{\sigma^{2}_{nar}}}\bigg) \\
    + f_{broad} \exp{\bigg(-\frac{1}{2}\frac{(\lambda - \lambda_{0})}{\sigma^{2}_{broad}}}\bigg)
\end{split}
\end{equation}

\noindent where $f_{c}$ is the continuum flux, which is assumed to be constant, $f_{nar}$ and $f_{broad}$ are the narrow and broad line amplitudes, and $\sigma^{2}_{nar}$ and $\sigma^{2}_{broad}$ are the individual line widths. Both Gaussians share a common line center $\lambda_{0}$.  
\par In the \Ha\ line model, we include additional Gaussian components for the \NII$\lambda\lambda$6550, 6585 doublet around the \Ha\ line. The intrinsic ratio of the line fluxes of the \NII\ lines are fixed to 1:2.94 \citep{Osterbrock2006}, while their linewidths and line centers are constrained to the narrow \Ha\ line.

\begin{figure*}[t!]
    \centering
    \includegraphics[width =\textwidth]{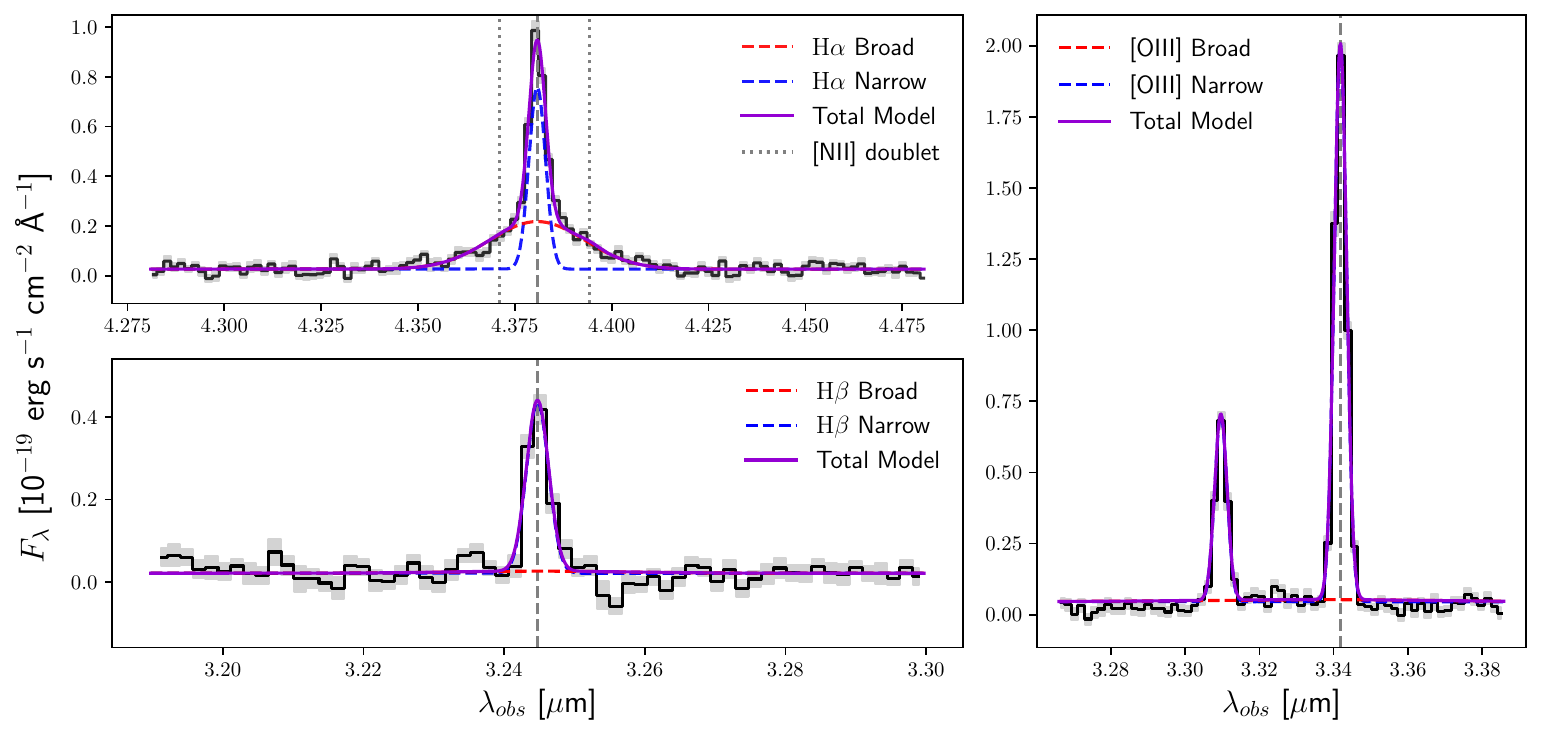}
    \caption{Observed-frame spectra (black histograms) and associated uncertainties (gray error bars) for RUBIES-EGS-47509. The top left panel shows the \Ha\ spectral region, the bottom left panel the \Hb\ spectra region, and the right panel the [OIII] doublet. Best-fit Gaussians are shown in dark blue for the narrow emission lines and red for the broad components. The total (narrow + broad) best fit model is shown in purple. Broad \Ha\ is significantly detected in this source ($>46\sigma$) but broad \Hb\ is not. The completed figure set (29 images) includes similar spectra for all BLAGN in our sample.}
    \label{fig:stacked spectra}
\end{figure*}


\par 
To fit the \Ha\ emission lines, we run \texttt{emcee} using 16 walkers and 25,000 steps
and implement a ``burn-in" of 10,000 steps, which are discarded from our final chains. We visually inspect the walker chains to confirm that these parameters are sufficient for robust and converged MCMC fits.  In our fits, we use flat priors for each model parameter. Additionally, we implement the following constraints on the fitted parameters:
\begin{enumerate}
    \item $f_{nar} > 0$ and $f_{broad} > 0$
    \item $\rm{FWHM}_\textit{{nar}} < 1000$ $\rm{km}$ $\rm{s^{-1}}$
    \item $\rm{FWHM}_\textit{{broad}} > 1000$ $\rm{km}$ $\rm{s^{-1}}$
\end{enumerate}
and $\lambda_{0}$ is allowed to vary within $250$ 
$\rm{km}$ $\rm{s^{-1}}$ of the reported literature redshift values. We confirmed that our best-fit broad and narrow line widths are statistically consistent with the values reported by \cite{Harikane2023, Maiolino2023, Kocevski2023} and \cite{Taylor2024}.

\par The above methodology is used to fit \Ha\ and we discuss our \Hb\ fitting procedure as follows. For the first run of our \Hb\ fits, we allow the line width to explore the parameter space indicated above. If the broad component of \Hb\ is not detected ($f_{broad} >3\sigma$), we constrain $\rm{FWHM}_\textit{{broad}}$ to within the measured $\rm{1\sigma}$ bounds of the \Ha\ $\rm{FWHM}_\textit{{broad}}$, and rerun the fit. Even after this iterative fitting process, broad \Hb\ is significantly detected ($>3\sigma$) in only 4/29 sources. The majority of our sample (25/29) exhibits weak broad \Hb\ emission despite being selected for significantly broad \Ha\ emission. 

\par Additionally, if \Hb\ $f_{narrow}$ is not detected, we constrain $\rm{FWHM}_\textit{{narrow}}$ to within $\rm{1\sigma}$ of \Ha\ $\rm{FWHM}_\textit{{narrow}}$. Narrow \Hb\ is detected in 24/29 sources. We report $3\sigma$ upper limits of broad \Hb\ emission and, when necessary, $3\sigma$ upper limits of narrow \Hb\ emission. Line fluxes for our sample are reported in Table \ref{Line Fluxes and FWHMs}.

\subsection{Outflows}
Large-scale outflows from the interstellar medium (ISM) can be observed as broad emission in the forbidden \OIII\ lines.  To confirm that our sample exhibits broad \Ha\ emission from an AGN and not from an outflow scenario, the \OIII $\lambda\lambda4959,5007$ doublet is checked for a broad component following the same fitting procedure described in \S\ref{Line Fitting}. The ratio of the [OIII] lines is fixed to 2.985:1 \citep{Storey2000}. 
 
 \par The vast majority of sources (25/29) with spectral coverage of the \OIII\ region do not have a significant broad \OIII\ component. Broad \OIII\ emission is observed in RUBIES-EGS-50052, also noted in \cite{Taylor2024}. We find a broad \Ha\ component with a $\rm{FWHM = 2051^{+31}_{-32}\,km\, s^{-1}}$ and an \OIII\ broad component with a $\rm{FWHM = 1030^{+17}_{-8}\,km\, s^{-1}}$. The \OIII\ broad component exhibits a significantly different velocity profile than the \Ha\ line, indicative that the origin is a galactic outflow and the system still hosts an AGN. We also detect weak ($<4\sigma$) \OIII\ broad emission in RUBIES-EGS-17416 with a $\rm{FWHM = 4943^{+299}_{-435}~km\, s^{-1}}$.  Visual inspection of this spectrum indicates that it appears consistent with noise associated with continuum emission. 
 
\par Spectra for the \OIII\ doublet region is not available for JADES-GS 8083 and JADES-GN 53757; \cite{Maiolino2023} rules out the outflow scenario for these two sources due to the symmetric nature of the broad \Ha\ emission line.  

\subsection{Stacked Spectra}\label{Stacked Spectra}
We produce median-stacked spectra, on both the \Ha\ and \Hb\ lines, with the sources that show no detection of the broad \Hb\ component. In total, this is 22/29 of the sources in our sample. We have confirmed with mock spectra, implementing the same narrow+broad Gaussian model as our fitting routine, that the mean broad-line flux is recovered in the stacked spectrum. Median-stacked spectra let us further explore the differences in the narrow and broad \Ha/\Hb\ ratio for BLAGN by increasing our signal-to-noise ratio. To stack our broad \Hb\ non-detections, we interpolate our spectra to a common velocity grid of $\sim$60~km~s$^{-1}$ per pixel and spanning $\pm 5000~\rm{km}~\rm{s^{-1}}$ around the \Ha\ and \Hb\ lines, that matches the lowest resolution spectrum in our sample (CEERS-11728, $z = 3.869$). Each spectrum is normalized by the peak of the \Ha\ line flux. We then combine the spectra by taking the median of the fluxes at each velocity pixel position in the velocity grid. Our median-stacked spectra result is shown in Figure \ref{fig:stacked spectra}. 

\par To estimate the errors for our stack, we follow the same stacking procedure as for the flux but median-stack the error spectrum for each source, and divide each error pixel by $\sqrt{N}$, where N is the number of sources in the stack. We then compare the normalized absolute median deviation (NMAD) of the continuum region around \Ha\ and \Hb\ to the median of the flux error over the spectral range around the emission lines. We found in both the \Ha\ and \Hb\ region that the errors are slightly overestimated, with NMAD($C_{f}$)/median($\sigma_{f}$) $ = 0.91$ and NMAD($C_{f}$)/median($\sigma_{f}$) $ = 0.60$ respectively. We elected to use the larger median-stacked errors as a conservative approach.

\begin{figure*}
    \centering
    \includegraphics[width =\textwidth]{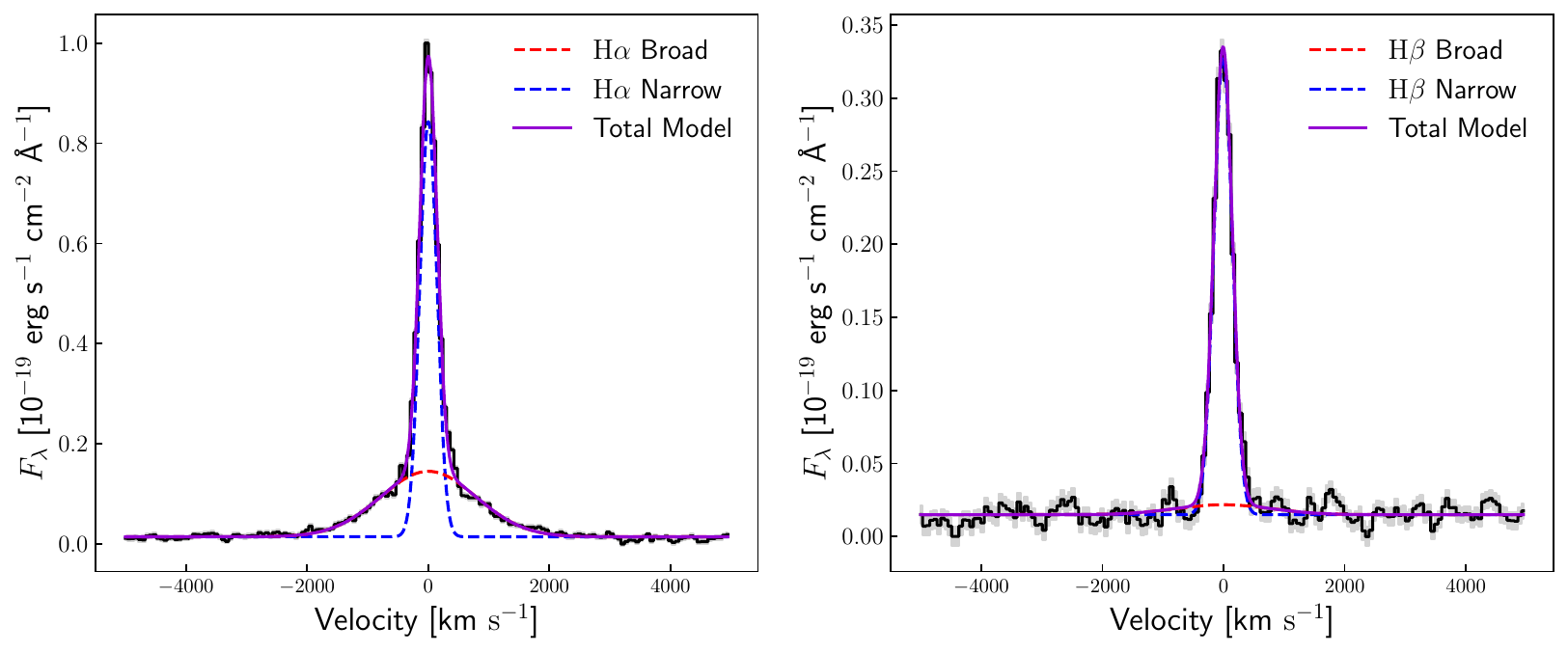}
    \caption{Median-stacked \Ha\ (left) and \Hb\ (right) emission lines for objects in our sample that exhibit no broad \Hb\ emission (25/29 sources). Our narrow line fits are shown in blue, broad line fits shown in red, and the total dual-Guassian fit is shown in purple. After stacking, broad \Hb\ emission is still undetected, suggesting that the mean population of AGN show significant dust attenuation in the BLR. The full stacking procedure is described in \S\ref{Stacked Spectra}. }
    \label{fig:stacked spectra}
\end{figure*}


\section{Results}\label{results}
The key result of this work is that even after median-stacking the majority of our sources (25/29), broad \Hb\ emission still remains undetected ($<3\sigma$).

\par The narrow and broad \Ha\ and \Hb\ emission line fluxes and corresponding $\rm{FWHM}_\textit{{broad}}$ for the individual sources in our sample are reported in Table \ref{Line Fluxes and FWHMs} and upper limits on \Hb\ are reported when necessary. The vast majority of our sources (25/29) do not have detected broad \Hb\ line emission despite having significantly ($>$5$\sigma$) detected broad \Ha\ emission. Additionally, 5 of the 29 sources have no ($<$3$\sigma$) detected \Hb\ emission at all (broad or narrow).

\par To estimate dust attenuation from the measured Balmer decrement, we follow the prescription provided by \cite{Momcheva2013}. We assume a \cite{Calzetti1997} attenuation curve and an intrinsic Balmer line ratio of $(\Ha/\Hb)_\mathrm{{int}} = 3.1$, the latter of which is appropriate for AGN BLR and NLR gas conditions \citep{Osterbrock2006}. When broad \Hb\ or narrow+broad \Hb\ is undetected, we determine the $3\sigma$ and $5\sigma$ lower limit on the \Ha/\Hb\ ratio. The full list of narrow and broad Balmer decrements (including both detections and limits) is found in Table \ref{Av measurements}. Emission lines associated with star-forming regions instead have an intrinsic line ratio of $(\Ha/\Hb)_\mathrm{{int}} = 2.86$ (for $n_e=10^2$~cm$^{-3}$, $T_e=10^4$~K, case B recombination), which would lead to slightly larger dust attenuation estimates than our assumption of $(\Ha/\Hb)_\mathrm{{int}} = 3.1$ for AGN-ionized gas. This choice in intrinsic ratio does not affect the general conclusions of this study.

\par
\begin{figure*}[t!]
    \centering
    \includegraphics[width = \textwidth]{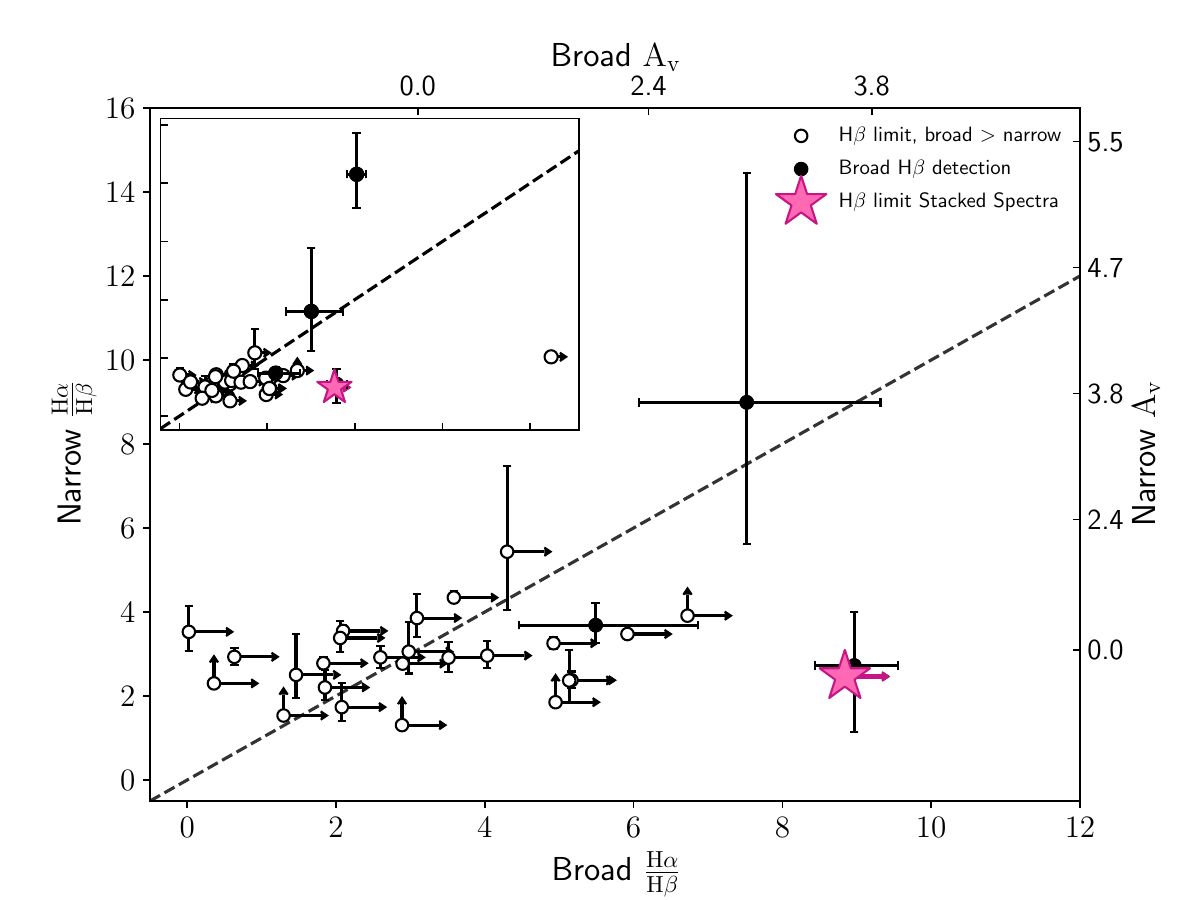}
    \caption{Narrow \Ha/\Hb\ vs. Broad \Ha/\Hb\ for the sample of 29 BLAGN. Filled circles represent sources with detected broad \Hb\ emission, while the open circles are galaxies with undetected broad \Hb\ and are plotted using the $3\sigma$ upper limit of broad \Hb. The pink star represents the stack of sources with no detected broad \Hb\ and is plotted using the $3\sigma$ upper limit of broad \Hb. The one-to-one line is shown by the dashed black line. An inset is included in the top left to show two extremely reddened sources, RUBIES-EGS-42406 (black circle) with detected broad \Hb\ emission and JADES 73488 (open circle). The $A_{v}$ for the broad and narrow flux ratios, calculated with the \cite{Calzetti1997} dust law and described in section \ref{results}, are shown on the dual x and y axes, respectively. 12/29 sources fall firmly below the one-to-one line, indicative of broad \Ha/\Hb\ $>$ narrow \Ha/\Hb.}
    \label{fig: narrow vs broad balmer decs}
\end{figure*}

 Figure \ref{fig: narrow vs broad balmer decs} shows the measured Balmer decrements and inferred V-band dust attenuation for the 29 AGN in our sample. The majority (26/29) of our sources have narrow-line \Ha/\Hb\ ratios that are consistent with the intrinsic value and little-to-no attenuation. The lower limits in broad \Hb\ are consistent with a wide range of broad-line attenuation; these sources are shown as open symbols in Figure \ref{fig: narrow vs broad balmer decs}.

\par Three sources, JADES-GN-73488, RUBIES-EGS-42046, and RUBIES-EGS-60935, show significant narrow line and broad line attenuation. Our most extremely attenuated source has a broad-line $A_{v} > 6.70$. This suggests an intrinsically high emission-line strength and we interpret the dustiest sources as unusual systems that are not representative of the bulk of the $z>3.5$ BLAGN sample \citep{Davis2023}. Specific narrow-line and broad-line Balmer ratios are given in Table \ref{Av measurements}.

\par JADES-GN-954 exhibits both broad \Ha\ and \Hb\ emission and has greater dust attenuation in the BL region. Additionally, RUBIES-EGS-60935 has detected, but not strong ($>2\sigma$) broad \Hb\ emission and shows similar attenuation in the narrow and broad region. 9 individual sources show broad \Ha/\Hb\ $>$ narrow \Ha/\Hb; \cite{Killi2023} also finds much higher attenuation in the BLR for a single source ($z \sim 4.53$). Sources in our sample with \Ha/\Hb\ $<$ narrow \Ha/\Hb\, falling left of the one-to-one line in Figure \ref{fig: narrow vs broad balmer decs}, generally have lower signal-to-noise ratios than sources right of the line. We expect deeper observations of these objects to show the result shown by the stack, broad \Ha/\Hb\ $>$ narrow \Ha/\Hb. 


\par In addition to the individual sources, we measure the narrow and broad Balmer decrement for our stacked spectra. We find narrow \Ha/\Hb\ = $2.47^{+0.05}_{-0.05}$ and broad \Ha/\Hb\ $> 8.85$. The narrow-line measurement is consistent with no attenuation and the broad-line ratio implies $A_{v} > 3.63$ for a \cite{Calzetti1997} attenuation curve. The median-stacked spectra is shown as a pink star in Figure \ref{fig: narrow vs broad balmer decs}. For an SMC attenuation curve \citep{Prevot1984, Bouchet1985}, we find a broad-line $A_{v} > 2.67$.

\begin{deluxetable*}{ccccccc}
\tablecaption{AGN Derived Properties}
\tablehead{\colhead{Name} & \colhead{$F_{\Ha,\mathrm{narrow}}$} & \colhead{$F_{\Ha,\mathrm{broad}}$} & \colhead{$\rm{FWHM}_{\Ha, \mathrm{broad}}$} & \colhead{$F_{\Hb,\mathrm{narrow}}$} &  \colhead{$F_{\Hb,\mathrm{broad}}$} & \colhead{$\rm{FWHM}_{\Hb, \mathrm{broad}}$} \\ \colhead{} & \multicolumn{2}{c}{{$[10^{-19} \rm{erg\ s^{-1} cm^{-2}}]$}} & \colhead{$[\mathrm{km ~s^{-1}}]$} & \multicolumn{2}{c}{{$[10^{-19} \rm{erg\ s^{-1} cm^{-2}}]$}} & \colhead{$[\mathrm{km ~s^{-1}]}$}}
\label{Line Fluxes and FWHMs}

\startdata
CEERS-11728 & $96.36_{-2.52}^{+2.61}$ & $22.82_{-2.62}^{+2.61}$ & $1067_{-  21}^{+40}$& $34.70_{-1.59}^{+1.57}$  & $<8.12$ & $-$\\
JADES-GN-73488 & $113.62_{-3.41}^{+3.46}$ & $227.91_{-3.98}^{+3.85}$ & $2098_{-  22}^{+21}$ & $22.33_{-1.18}^{+1.22}$ & $<7.81$ &  $-$\\
JADES-GN-11836 & $157.16_{-2.49}^{+2.40}$ & $39.69_{-3.44}^{+3.45}$ & $1617_{-  59}^{+62}$ & $44.29_{-1.65}^{+1.64}$  & $<12.46$ & $-$ \\
JADES-GN-53757 & $41.43_{-3.65}^{+3.75}$ & $72.73_{-4.44}^{+4.49}$ & $1916_{-  72}^{+87}$ & $13.55_{-2.34}^{+2.41}$ & $<16.23$ & $-$ \\
CEERS-1665 & $453.40_{-7.21}^{+6.70}$ & $107.85_{-7.58}^{+7.65}$ & $1722_{-  56}^{+59}$ & $104.33_{-3.07}^{+3.10}$ & $<18.55$ & $-$\\
CEERS-1236 & $12.01_{-1.19}^{+1.82}$ & $28.30_{-3.22}^{+3.29}$ & $3290_{- 143}^{+155}$  & $7.05_{-1.19}^{+1.26}$ & $<7.72$ &  $-$ \\
JADES-GS-8083  & $84.20_{-1.35}^{+1.33}$ & $32.54_{-1.96}^{+2.00}$ &  $1728_{-  52}^{+55}$ & $25.86_{-0.97}^{+0.98}$ & $<4.25$ &  $-$ \\
RUBIES-EGS-46985 & $89.19_{-3.19}^{+3.55}$ & $61.79_{-3.40}^{+3.44}$ &  $1052_{-  17}^{+34}$ & $26.70_{-2.83}^{+2.89}$ & $<22.17$ & $-$  \\
RUBIES-EGS-17416 & $24.68_{-1.49}^{+1.58}$ & $25.80_{-1.51}^{+1.44}$ &  $1078_{-  23}^{+39}$ & $11.32_{-1.84}^{+1.71}$  & $<10.29$ &   $-$\\
JADES-GN-62309 & $33.98_{-1.63}^{+1.63}$ & $12.05_{-2.47}^{+2.39}$ & $1076_{-  40}^{+70}$ & $9.63_{-1.37}^{+1.35}$ & $<12.64$ &  $-$\\
RUBIES-EGS-17301 & $37.98_{-3.55}^{+3.68}$ & $54.97_{-8.57}^{+9.51}$ & $2044_{- 170}^{+253}$ & $<3.77$ & $<21.44$ &  $-$\\
JADES-GN-77652 & $27.34_{-3.01}^{+3.47}$ & $33.51_{-4.17}^{+3.91}$ & $1026_{-  27}^{+49}$ & $10.96_{-2.88}^{+2.87}$ & $<11.22$ &  $-$\\
RUBIES-EGS-50052 & $316.36_{-2.45}^{+2.43}$ & $257.55_{-3.18}^{+3.22}$ & $2052_{-  14}^{+13}$ & $91.02_{-1.75}^{+1.75}$ & $<38.08$ &  $-$\\
RUBIES-EGS-13872 & $15.61_{-1.23}^{+1.11}$ & $11.28_{-1.44}^{+1.41}$  &  $1374_{-  99}^{+128}$  & $<4.57$ & $<2.23$ &  $-$\\
RUBIES-EGS-42046 & $635.30_{-12.81}^{+12.49}$ & $2026.03_{-20.20}^{+20.22}$ & $3465_{-  17}^{+17}$ & $30.32_{-4.25}^{+4.31}$ & $199.12_{-9.41}^{+9.62}$ &  $2526_{-  61}^{+66}$ \\
RUBIES-EGS-60935 & $77.49_{-4.67}^{+4.77}$ & $157.06_{-5.21}^{+5.06}$  & $2016_{-  36}^{+38}$ & $8.75_{-3.73}^{+5.71}$ & $21.55_{-4.20}^{+4.23}$ &   $1149_{- 116}^{+513}$ \\
RUBIES-EGS-926125 & $24.72_{-1.06}^{+1.07}$ & $53.48_{-1.65}^{+1.62}$ & $1586_{-  25}^{+25}$ & $8.35_{-0.83}^{+0.85}$ & $<9.24$ &   $-$\\
CEERS-746    & $15.34_{-2.69}^{+2.70}$ & $48.06_{-2.89}^{+2.83}$ & $1609_{-  71}^{+75}$ & $<0.11$ & $<7.21$ &  $-$\\
JADES-GN-1093  & $25.47_{-1.31}^{+1.31}$ & $14.25_{-3.16}^{+3.53}$ & $2590_{- 254}^{+329}$ & $<3.31$ & $<9.71$ &  $-$\\
RUBIES-UDS-29813 & $54.13_{-2.26}^{+2.30}$ & $82.79_{-5.28}^{+5.25}$ & $2809_{- 111}^{+117}$ & $<4.34$ & $<6.72$ &  $-$\\
RUBIES-UDS-19521 & $31.84_{-2.36}^{+2.46}$ & $28.96_{-3.69}^{+4.59}$ & $1895_{- 168}^{+380}$ & $10.95_{-1.03}^{+1.03}$ & $<5.12$ &  $-$\\
RUBIES-UDS-47509 & $39.29_{-1.94}^{+1.98}$ & $59.02_{-2.63}^{+2.69}$  & $1972_{-  54}^{+58}$ & $16.48_{-1.09}^{+1.06}$ & $<7.87$ &  $-$\\
RUBIES-EGS-27915 & $165.85_{-4.65}^{+4.74}$ & $60.93_{-9.95}^{+9.82}$ & $2795_{- 118}^{+130}$ & $56.57_{-3.45}^{+3.50}$ & $<47.37$ &  $-$\\
JADES-GN 61888 & $32.23_{-2.02}^{+1.99}$ & $33.29_{-2.58}^{+2.57}$ & $1332_{-  56}^{+58}$ & $8.35_{-0.98}^{+0.94}$  & $<7.84$ &  $-$\\
JADES-GS 10013704 & $23.00_{-1.04}^{+1.05}$ & $40.35_{-2.74}^{+2.72}$ &  $2513_{-  71}^{+75}$ & $7.88_{-0.64}^{+0.65}$ & $<12.10$ &  $-$\\
CEERS 00397    & $177.50_{-3.17}^{+3.03}$ & $33.49_{-4.76}^{+4.79}$ & $1946_{- 115}^{+117}$ & $64.08_{-1.54}^{+1.53}$ & $<6.77$ & $-$ \\
RUBIES-EGS-49140 & $207.33_{-22.72}^{+22.19}$ & $946.32_{-23.08}^{+21.40}$ & $2932_{-  24}^{+26}$ & $56.25_{-16.44}^{+61.21}$ & $107.72_{-6.43}^{+6.26}$ &  $2254_{-  53}^{+58}$\\
JADES-GN 954   & $53.28_{-3.74}^{+3.92}$ & $159.66_{-8.22}^{+8.71}$ & $1808_{-  49}^{+51}$ & $14.45_{-1.58}^{+1.60}$ & $828.24_{-115.03}^{+114.84}$ &   $2010_{- 154}^{+185}$\\
RUBIES-UDS-807469 & $10.32_{-1.98}^{+2.73}$ &  $48.75_{-2.72}^{+2.76}$ & $1603_{-  53}^{+64}$ & $4.40_{-0.61}^{+0.64}$ & $<7.27$ & $-$\\ 
\enddata
\tablecomments{We do not report broad \Hb\ FWHMs for the sources with non-detections, these are denoted with a ``$-$''.}
\end{deluxetable*}

\section{Discussion and Conclusions}
In this work, we present dust attenuation measurements for 29 broad-line identified AGN in the JWST deep fields. We also further investigate the optical dust attenuation of these sources using a stacking analysis of sources that have no significantly detected broad \Hb\ emission. The non-detection of \Hb\ and significantly detected \Ha\ ($>$15$\sigma$) after stacking indicates that these sources are, on average, heavily dust-attenuated in the rest-frame optical. Narrow \Ha/\Hb\ = $2.55^{+0.07}_{-0.07}$ (consistent with zero dust attenuation\footnote{The stacked narrow \Ha/\Hb\ is significantly lower than the fiducial value of $\Ha/\Hb = 2.86$. This likely implies higher electron temperature for the gas (for example, $T_e = 2 \times 10^4$~K results in $\Ha/\Hb = 2.75$ for case B recombination), consistent with previous work studying the ISM conditions of high-redshift galaxies \citep[e.g.,][]{Trump2023,Backhaus2024,Sanders2024}})
and broad \Ha/\Hb\ $> 8.85$  ($A_V > 3.63$) for the stack suggests that the origin of the narrow and broad components of the emission lines originate from different regions in the galaxy. The broad emission lines are emitted from near the AGN and their high attenuation implies that the red optical colors of LRDs are driven by dust-attenuated emission from the central AGN. Meanwhile, the narrow lines are presumably emitted from extended scales, and the low attenuation of the narrow lines implies that the blue/UV colors of LRDs are dominated by star formation processes in the host galaxy. 

\par The narrow Balmer lines in the majority of our sources (26/29) are consistent with little to no dust attenuation. Our distribution of narrow-line Balmer decrements is consistent with \cite{Shapley2023} measurements for a population of non-BL AGN. \cite{Sandles2023} also finds consistently low Balmer decrements, where the median is $2.88 \pm0.08$ for a sample of 51 galaxies. The narrow line Balmer decrement of the stacked spectra also shows no dust attenuation. The central AGN in these sources are consistently much more reddened than what is seen on the galactic scale, further suggesting that the blue UV colors seen in the SEDs is from star-formation in the galaxy.

\par The source of attenuation around the central AGN could be explained by a dusty torus of obscuring gas near the AGN \citep{Krolik1988}. However, mid-infrared observations of LRDs are inconsistent with the dusty torus model \citep{Williams2024, PerezGonzalez2024}. Rather than a canonical torus, the AGN may be attenuated by larger-scale dust in the galaxy nucleus that is somewhat cooler than a torus but is still much more compact than the host galaxy starlight \citep{Buchner2017}. The observed dust attenuation may also be described by a polar dust model \citep{Yang2020, Buat2021} that is more compact than the narrow-line region but colder than the torus.

\par Higher attenuation around the AGN is also consistent with the lack of X-ray detections for these sources \citep{Maiolino2024}. Two of our sources, RUBIES-EGS-42046 and RUBIES-EGS-49140, show strong Balmer absorption lines, which requires extremely high densities of neutral hydrogen gas \citep{Hall2007, Inayoshi2024}. High-density absorbing material around the AGN may dampen X-ray emission and explain the non-detections of the LRDs in deep X-ray fields, even after a stacking analysis. Additionally, our results show a dramatic bimodality of the broad and narrow-line attenuation for high-redshift AGN, which suggests that these regions in the galaxy are physically distinct. The dramatically different attenuation of the broad and narrow lines disfavors non-AGN models \citep{Kokubo2024, Baggen2024} due to gas that is more likely to have continuous velocity and attenuation distributions. 


\par The large population of broad-line AGN and LRDs at high-redshifts is one of the biggest surprised revealed by JWST. This work shows, on average, high-redshift broad-line AGN have highly attenuated nuclei but little-to-no attenuation affecting the extended host galaxy. However, currently available JWST spectroscopy is insufficient to resolve the nature of the dust in individual sources and to study the distribution of dust attenuation as attenuation as a function of galaxy and AGN properties. This population represents a look into obscured black hole formation growth in early epochs and is important to fully understanding black hole-galaxy coevolution. Deeper and/or NIRSpec IFU spectroscopy is needed to further disentangle the obscured nature of AGN and host galaxy emission and growth at cosmic dawn.

\par We thank the JADES and RUBIES team for their effort designing and executing their programs and for making the data publicly available. We acknowledge the work of our colleagues in the CEERS collaboration and everyone involved in the JWST mission.
MB, JRT, and KD acknowledge support from NASA grants JWST-ERS-01345, JWST-AR-01721, and NSF grant CAREER-1945546.

\begin{deluxetable*}{ccccc}
\tablecaption{Narrow and Broad Reddening of BL AGN}\label{Av measurements}
\tablewidth{20pt}

\tablehead{\colhead{Name} & \colhead{Narrow \Ha/\Hb} &  \colhead{Narrow $A_{v}$} & \colhead{Broad \Ha/\Hb} & \colhead{Broad $A_{v}$}} 
\startdata
CEERS-11728 & $2.78_{-0.14}^{+0.15}$ & $\simeq$ 0 & $>\!\!1.83\,(>\!\!1.83)$ & $\simeq$ 0 \\
JADES-GN-73488 & $5.09_{-0.29}^{+0.32}$ & $1.72_{-0.21}^{+0.21}$ & $>\!\!21.48\,(>\!\!21.20)$ & $>\!\!6.70\,(>\!\!6.66)$ \\
JADES-GN-11836 & $3.55_{-0.14}^{+0.14}$ & $0.47_{-0.14}^{+0.14}$ & $>\!\!2.14\,(>\!\!2.10)$ & $\simeq$ 0 \\
JADES-GN-53757 & $3.06_{-0.52}^{+0.71}$ & $\simeq$ 0 & $>\!\!3.02\,(>\!\!2.98)$ & $\simeq$ 0 \\
CEERS-1665 & $4.34_{-0.14}^{+0.15}$ & $1.17_{-0.11}^{+0.12}$ & $>\!\!3.84\,(\!\!>3.58)$ & $\simeq$ 0\\
CEERS-1236 & $1.74_{-0.33}^{+0.57}$ & $\simeq$ 0 & $>\!\!2.11\,(>\!\!2.08)$ & $\simeq$ 0 \\
JADES-GS-8083 & $3.26_{-0.13}^{+0.14}$ & $0.17_{-0.14}^{+0.15}$ & $>\!\!5.11\,(>\!\!4.92)$ & $>\!\!1.73\,(>\!\!1.60)$ \\
RUBIES-EGS-46985 & $3.35_{-0.35}^{+0.41}$ & $0.26_{-0.38}^{+0.40}$ & $>\!\!1.97\,(>\!\!1.85)$ & $\simeq$ 0 \\
RUBIES-EGS-17416 & $2.18_{-0.30}^{+0.44}$ & $\simeq$ 0 & $>\!\!1.76\,(>\!\!1.69)$ & $\simeq$ 0 \\
JADES-GN-62309 & $3.53_{-0.45}^{+0.61}$ & $0.45_{-0.48}^{+0.55}$ & $>\!\!0.25\,(>\!\!0.02)$ & $\simeq$ 0 \\
RUBIES-EGS-17301 & $>\!\!1.45\,(>\!\!1.44)$ & $\simeq$ 0 & $>\!\!1.49\,(>\!\!1.49)$ & $\simeq$ 0 \\
JADES-GN-77652 & $2.50_{-0.56}^{+0.96}$ & $\simeq$ 0 & $>\!\!1.55\,(>\!\!1.46)$ & $\simeq$ 0 \\
RUBIES-EGS-50052 & $3.48_{-0.07}^{+0.07}$ & $0.40_{-0.07}^{+0.07}$ & $>\!\!6.19\,(>\!\!6.11)$ & $>\!\!2.39\,(>\!\!2.35)$ \\
RUBIES-EGS-13872 & $>\!\!1.38\,(>\!\!1.36)$ & $\simeq$ 0 & $>\!\!2.73\,(>\!\!2.73)$ & $\simeq$ 0 \\
RUBIES-EGS-42046 & $20.95_{-2.65}^{+3.47}$ & $6.62_{-0.47}^{+0.53}$ & $10.18_{-0.47}^{+0.51}$ & $4.12_{-0.16}^{+0.17}$ \\
RUBIES-EGS-60935 & $8.85_{-3.47}^{+6.52}$ & $3.63_{-1.72}^{+1.91}$ & $7.29_{-1.22}^{+1.77}$ & $2.96_{-0.63}^{+0.75}$ \\
RUBIES-EGS-926125 & $2.96_{-0.30}^{+0.35}$ & $\simeq$ 0 & $>\!\!4.67\,(>\!\!4.62)$ & $>\!\!1.42\,(>\!\!1.38)$ \\
CEERS-746 & $>\!\!1.85\,(>\!\!1.80)$ & $\simeq$ 0 & $>\!\!4.99\,(>\!\!4.95)$ & $>\!\!1.65\,(>\!\!1.62)$ \\
JADES-GN 1093 & $>\!\!2.30\,(>\!\!2.27)$ & $\simeq$ 0 & $>\!\!0.44\,(>\!\!0.36)$ & $\simeq$ 0 \\
RUBIES-UDS-29813 & $>\!\!3.91\,(>\!\!2.98)$& $>\!\!0.81\,(>\!\!0.80)$ & $>\!\!7.36\,(>\!\!6.73)$ &  $\simeq$ 0 \\
RUBIES-UDS-19521 & $2.91_{-0.33}^{+0.37}$ & $\simeq$ 0 & $>\!\!3.55\,(>\!\!3.51)$ &  $\simeq$ 0 \\
RUBIES-UDS-47509 & $2.38_{-0.19}^{+0.20}$ & $\simeq$ 0 & $>\!\!5.15\,(>\!\!5.13)$ & $>\!\!1.76\,(>\!\!1.74)$ \\
RUBIES-EGS-27915 & $2.93_{-0.19}^{+0.21}$ & $\simeq$ 0 & $>\!\!0.69\,(>\!\!0.63)$ & $\simeq$ 0\\
JADES-GN-61888 & $3.86_{-0.45}^{+0.57}$ & $0.75_{-0.43}^{+0.48}$ & $>\!\!3.11\,(>\!\!3.09)$ & $>\!\!0.01\,(0)$ \\
JADES-GS-10013704 & $2.92_{-0.25}^{+0.28}$ & $\simeq$ 0 & $>\!\!2.63\,(>\!\!2.60)$ & $\simeq$ 0 \\
CEERS-397 & $2.77^{+0.84}_{-0.08}$ & $\simeq$ 0 & $>\!\!2.84\,(>\!\!2.82)$ & $\simeq$ 0 \\
RUBIES-EGS-49140 & $2.73_{-1.59}^{+1.28}$ & $\simeq$ 0 & $8.97_{-0.526}^{+0.593}$ & $3.68_{-0.209}^{+0.222}$ \\
JADES-GN-954 & $3.69_{-0.43}^{+0.52}$ & $0.60_{-0.43}^{+0.46}$ & $5.49_{-1.03}^{+1.37}$ & $1.98_{-0.72}^{+0.77}$ \\
RUBIES-UDS-807469 & $2.37_{-0.54}^{+0.72}$ & $\simeq$ 0 & $>\!\!5.19\,(>\!\!5.13)$ & $>\!\!1.79\,(>\!\!1.75)$\\
\enddata

\tablecomments{Here we are reporting the $3\sigma$ lower limit on the Balmer decrement measurements when necessary. The $5\sigma$ limit is additionally given in the parentheses.}
\end{deluxetable*}

\facilities{\textit{JWST}}
\software{astropy: \cite{astropy2013,astropy2018,Astropy2022}, scipy: \cite{scipy2020}, emcee: \citep{emcee}, numpy: \cite{numpy}}

\newpage 

\bibliographystyle{aasjournal}
\bibliography{citations}

\end{document}